\documentclass[10pt,twocolumn]{article}

\usepackage{times}
\usepackage{fullpage}
\usepackage{amssymb}
\usepackage{amsmath}
\usepackage{paralist, tabularx}

\input{styling/codestyling.sty}
\input{styling/sizing.sty}
\input{styling/macros.sty}
\usepackage{styling/usenix}

\begin{document}

\pagenumbering{gobble}

\title{\Large\bf Incremental Consistency Guarantees for Replicated Objects}

\author{Rachid Guerraoui}
\author{Matej Pavlovic}
\author{Dragos-Adrian Seredinschi\footnote{Authors appear in alphabetical order.}}
\affil{School of Computer and Communication Sciences,\\
\'{E}́cole Polytechnique F\'{e}́d\'{e}́rale de Lausanne (EPFL), Switzerland\\
\{rachid.guerraoui, matej.pavlovic, dragos-adrian.seredinschi\}@epfl.ch}
\date{}

\maketitle

%!TEX root = ../bundle.tex

\begin{abstract}

%% Programming with replicated objects is difficult, as developers must face the fundamental trade-off between consistency and performance head on.
%% \name\ are an interface between applications and their complex replication stacks.
%% % The main goal of this abstraction is to simplify programming with replicated data.
%% \name\ hide are based on abstract consistency levels, hiding system-specific protocols, and enabling application code to be portable and simpler.
%% A \interface\ is a client-side container for a replicated object, which exposes multiple views of this object, in incremental consistency (and latency) steps.
%% Applications can thus use incremental guarantees on an object, and traverse the consistency/performance trade-off at runtime.

%% Programming with replicated objects is difficult.
%% Developers must face the fundamental trade-off between consistency and performance head on, while struggling with the complexity of underlying storage stacks.
%% \emph{\name} are a novel abstraction that hides most of this complexity, allowing developers to focus on the task of balancing consistency and performance.
%% To aid developers with this task, \name\ capture an incremental refinement on the consistency guarantees for an ongoing operation.
%% With little resource overhead, applications can benefit from both a fast (possibly inconsistent) result, and a consistent (correct) result that arrives later.

Programming with replicated objects is difficult.
Developers must face the fundamental trade-off between consistency and performance head on, while struggling with the complexity of distributed storage stacks.
We introduce \emph{\name}, a novel abstraction that hides most of this complexity, allowing developers to focus on the task of balancing consistency and performance.
To aid developers with this task, \name\ provide \emph{incremental consistency guarantees}, which capture successive refinements on the result of an ongoing operation on a replicated object.
In short, applications receive both a preliminary---fast, possibly inconsistent---result, as well as a final---consistent---result that arrives later.

% With little resource overhead, applications can benefit from both a fast (possibly inconsistent) result, and a consistent (correct) result that arrives later.

% We implement three applications using \name : (1) a notification system backend (on top of Cassandra), (2) a photo album application (on top of a causally-consistency system), and (3) a ticket selling system (on top of ZooKeeper).
% All applications interact with the underlying system via \interface s.
% Our evaluation on Amazon EC2 shows that \name\ can mask up to \needsrev{67\%} of the latency of strongly consistent storage access at little cost (\needsrev{10\%} bandwidth increase, \needsrev{12\%} throughput drop) for low contention workloads.
% For high contention, the incorrect weakly consistent results (25\%) are amended by \name\ at a bandwidth increase of only 27\%.

% \emph{\name} are a novel abstraction that hides most of this complexity, allowing developers to focus on the task of balancing consistency and performance.
% To aid developers with this task, \name\ capture an \emph{incremental} refinement on the consistency guarantees for an ongoing operation.
% With little resource overhead, a

% towards improving performance applications can
We show how to leverage incremental consistency guarantees by speculating on preliminary values, trading throughput and bandwidth for improved latency.
We experiment with two popular storage systems (Cassandra and ZooKeeper) and three applications: a Twissandra-based microblogging service, an ad serving system, and a ticket selling system.
Our evaluation on the Amazon EC2 platform with YCSB workloads A, B, and C shows that we can reduce the latency of strongly consistent operations by up to $40\%$ (from $100ms$ to $60ms$) at little cost ($10$\% bandwidth increase, $6$\% throughput drop) in the ad system.
Even if the preliminary result is frequently inconsistent ($25$\% of accesses), incremental consistency incurs a bandwidth overhead of only $27$\%.

\end{abstract}

%!TEX root = ../bundle.tex

\section{Introduction}
\label{sec:introduction}

Replication is a crucial technique for achieving performance---i.e., high availability and low latency---in large-scale applications.
Traditionally, strong consistency protocols hide replication and ensure correctness by exposing a single-copy abstraction over replicated objects~\cite{Corbett:2013:Spanner,lam98paxos}.
There is a trade-off, however, between consistency and performance~\cite{abad12pacelc,brewer2012cap,gilbert2002brewer}.
Weak consistency~\cite{decandia2007dynamo} boosts performance,
% \needsrev{---possibly by orders of magnitude \cite{terry13pileus}---}
but introduces the possibility of incorrect (anomalous) behavior.
 % does so at the expense of possible

% The specific choice of a consistency protocol in an application thus hinges on one out of two basic needs---either for performance or for correctness.

% Given the trade-off between strong consistency and performance~\cite{abad12pacelc,brewer2012cap,gilbert2002brewer}, however,
% To ensure correctness, many applications use strong consistency protocols, which expose a single-copy abstraction over the replicated data~\cite{Corbett:2013:Spanner,lam98paxos}.
% As performance is often the overriding goal, developers prefer to avoid strong consistency, and employ it only when trully necessary.

% In practice, the performance gap between different consistency levels can also vary in different scenarios.
% In some cases, e.g. when a client is equally far from a primary and a backup replica of a replicated service, there is no reason for sacrificing strong consistency.

A common argument in favor of weak consistency is that such anomalous behavior is rare in practice.
Indeed, studies reveal that on expectation, weakly consistent values are often correct even with respect to strong consistency~\cite{bailis2012PBS,lu2015existential}.
Applications which primarily demand performance
% that put the main focus on performance
thus forsake stronger models and resort to weak consistency~\cite{ajouxchallenges,decandia2007dynamo}.

% Many applications put focus on performance, so given this observation, they forsake stronger models.

% In other cases, the performance difference can reach orders of magnitude~\cite{terry13pileus}, justifying the choice of weak consistency even for the price of occasional incorrect behavior of the application.
%\newpage %%% watch out
There are cases, however, where applications often diverge from correct behavior due to weak consistency.
As an extreme example, consider that an execution of YCSB workload A~\cite{co10ycsb} in Cassandra~\cite{la10cassandra} on a small $1K$ objects dataset reveals stale values for 25\% of weakly consistent read operations (\Cref{fig:divergence} in~ \Cref{sec:evaluation}).
This happens when using the \emph{Latest} distribution, where read activity is skewed towards popular items~\cite{co10ycsb}.
In other cases, even very rare anomalies are unacceptable (e.g. when handling sensitive data such as user passwords), making strongly consistent access a necessity.
For this class of applications, correctness supersedes performance, and strong consistency thus takes precedence~\cite{Corbett:2013:Spanner}.

%% \begin{figure}[t]
%% \centerline{\includegraphics[width=0.8\columnwidth]{./figures/divergence}}
%% \caption{YCSB benchmark: percentage of divergence in Cassandra between reads with $R=1$ and $R=2$.}
%% \label{fig:divergence}
%% \vspace{-.5cm}
%% \end{figure}

% In practice, another critical concern when choosing a consistency model is the various performance gaps across different models.
% Consider a client of a geo-replicated system which is equally apart from the primary and the backup replica.
% Typically, this client would experience no penalty for using strong consistency (reading from primary) compared to weak consitency (using the backup).
% When the performance gaps are large---as it may often happen in practice~\cite{terry13pileus}---many applications prefer weak consistency.

% In other words, when there is non-negligible divergence, applications cannot escape to weak consistency.
% Likewise, when there are sizable performance gaps between different models, strong consistency is relatively expensive and applications would pay a big price for it.
% In such a scenario, we say that applications are in a \emph{gray zone}: they operate in conditions where strong consistency is required (for at least some parts of the application), but it is very costly (see~\Cref{fig:gray-zone}).

\begin{figure}[b]
\vspace{-.7cm}
\centerline{\includegraphics[width=0.9\columnwidth]{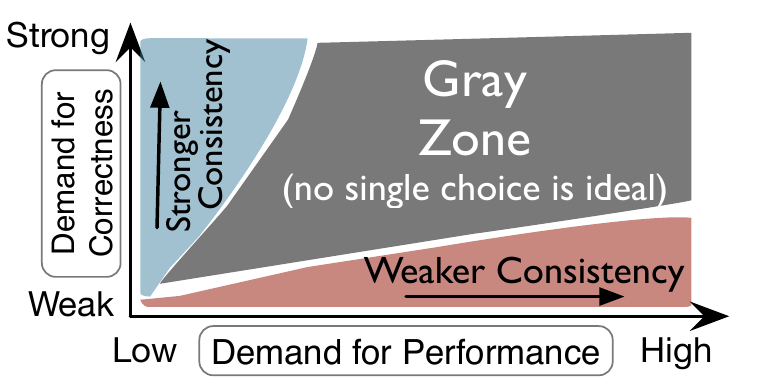}}
\caption{Many applications fall into a \emph{gray zone}, torn between the need for both performance and correctness.}
\label{fig:gray-zone}
\end{figure}

There is also a large class of applications which do not have a single, clear-cut goal (either performance or correctness).
Instead, such applications aim to satisfy \emph{both} of these conflicting demands.
These applications fall in a \emph{gray zone}, somewhere in-between the two previous classes, as we highlight in~\Cref{fig:gray-zone}.
Typically, these applications aim to strike an optimal balance of consistency and performance by employing different consistency models, often at the granularity of individual operations~\cite{bail15feral,cooper2008pnuts,kra09ration,li2012making,terry13pileus}.
Choosing the appropriate consistency model, even at this granularity, is hard, and the result is often sub-optimal, as developers still end up with fixing a certain side of the consistency/performance trade-off (and sacrificing the other side).

%% A notable problem with this approach is that developers are still dealing with separate consistency models, where each model has certain advantages and drawbacks.
%% Some operations exhibit high performance with weak consistency, while others run slower with stronger consistency.

% Programming in the gray area is difficult.
% Typically, developers strive to mix separate consistency models across operations.
% The goal is to strike an optimal balance of consistency and performance for each \emph{individual} operation.

Moreover, programming in the gray area is difficult, as developers have to juggle different consistency models in their applications~\cite{cooper2008pnuts,kra09ration}.
% Such fine-tuning is important, since there are large performance gaps---by orders of magnitude---across distinct consistency levels~\cite{terry13pileus}.
If programming with a single consistency model (such as weak consistency~\cite{Corbett:2013:Spanner}) is non-trivial, then mixing multiple models is even harder~\cite{li14automa}.
In their struggle to optimize performance with consistency, developers must go up against the full complexity of the underlying storage.
This includes choosing locations (cache or backup or primary replica), dealing with coherence and cache-bypassing, or selecting quorums.
These execution details reflect as a burden on developers, complicate application code, and lead to bugs~\cite{er13your,lu2015existential}.

Our goal is to help with the programming of applications located in the gray area.
%We start from the observation that no single consistency model is \emph{ideal}, providing both high performance and strong consistency (correctness) at the same time~\cite{abad12pacelc,gilbert2002brewer}.
We accept as a fact that no single consistency model is ideal, providing both high performance and strong consistency (correctness) at the same time~\cite{abad12pacelc,gilbert2002brewer}.
Our insight is to approach this ideal in complementary steps, by \emph{combining consistency models in a single operation}.
Briefly, developers can invoke an operation on a replicated object and obtain multiple, incremental \emph{views} on the result, at successive points in time.
Each view reflects the operation result under a particular consistency model.
Initial (preliminary) views deliver with low latency---but weak consistency---while stronger guarantees arrive later.
We call this approach \emph{incremental consistency guarantees} (\incguarantees ).

% Beside divergence, there is a second common metric which can play either in favor or against weak consistency: the performance gap between different consistency models.

%% In practice, as the rate of divergence increases, so does the chance of anomalous behavior, motivating the need for strong consistency.
%% Similarly, as the performance gap between different models widens, then weak consistency protocols become more pressing, in order to preserve performance.
%% It is unclear, however, what consistency model to use when these metrics do not agree: what is the ideal protocol when there is both a high divergence and large performance gap?

We introduce \name , an abstraction which grants developers a clean, consistency-based interface for accessing replicated objects, clearly separating semantics from execution details.
% drawing a clear boundary between semantics and execution details.
This abstraction reduces programmer effort by hiding storage-specific protocols, e.g., selecting quorums, locations, or managing coherence.
\interface s are based on \emph{Promises} \cite{lisk88promises}, which are placeholders for a single value that becomes available in the future.
\interface s generalize Promises by representing not a single, but multiple future values, corresponding to incremental views on a replicated object.

%% Traditional work on replicated protocols assumes a request-reply model to access replicated objects.
%% \name\ represent a departure from this practice.
To the best of our knowledge, our abstraction is the first which enables applications to build on \incguarantees .
As few as \emph{two} views suffice for \incguarantees\ to be useful.
The advantage of \incguarantees\ is that applications can speculate on the preliminary view, hiding the latency of strong consistency, and thereby improving performance~\cite{we09toler}.
Speculating on preliminary responses is expedient considering that, in many systems, weak consistency provides correct results on expectation~\cite{bailis2012PBS,lu2015existential}.

% \begin{figure}[t]
% \centerline{\includegraphics[width=0.75\columnwidth]{./figures/incremental-tradeoffs}}
% \caption{No single consistency model is ideal: \name\ allow applications to traverse models incrementally.}
% \label{fig:incremental-traversal}
% \vspace{-.5cm}
% \end{figure}

% \needsrev{Moreover, some of these operations (e.g., on sensitive data such as user accounts) require strong guarantees and incur high latency.}

% Speculation, in this scenario, would mean prefetching the dependent objects based on a preliminary (weakly-consistent) view.

Speculation with \incguarantees\ is applicable to a wide range of scenarios.
Consider, for instance, that a single application-level operation can aggregate multiple---up to hundreds of---storage-level objects~\cite{ajouxchallenges,dea13tailscale,li15domino,ma15thesis}.
Since these objects are often inter-dependent, they can not always be fetched in parallel.
With \incguarantees , the application can use the fast preliminary view to speculatively prefetch any dependent objects.
By the time the final (strongly consistent) view arrives, the prefetching would also finish.
If the preliminary result was correct (matching the final one), then the speculation is deemed successful, reducing the overall latency of this operation.

% Scenarios where speculation is appropriate include login operations (by prefetching the user's home page or timeline, for instance), as well as read operations which translate into a chain of lower-level reads, as is the case for shopping carts, friends lists, notification messages, and so on.

Alternatively, \incguarantees\ can open the door to exploiting application-specific semantics for optimizing performance.
Imagine an application requiring a monotonically increasing counter to reach some pre-defined threshold (e.g., number of purchased items in a shop required for a fidelity discount).
If a weakly consistent view of the counter already exceeds this threshold, the application can proceed without paying the latency price of a strongly consistent view.

%% \needsrev{Alternatively, applications can also leverage \incguarantees\ by actually \emph{exposing} preliminary data to users, and then revising the output when the final value arrives.
%% Indeed, this alternative use-case \needsrev{already happens in practice} in some applications: it provides to users the illusion of high responsiveness (i.e., that the service is live), eventually ensuring a correct output.}
% We note that side-effects are not a concern, as updates typically run off the critical path for the very reason of ensuring performance~\cite{dea13tailscale}.
% Such side-effect operations would thus execute asynchronously on the final result.
% We design our interface, however, to be general, so we do provide a programmatic way to abort a miss-speculations.

The high-level abstraction centered on consistency models, coupled with the performance benefits of enabling speculation via \incguarantees , are the central contributions of \name .
We evaluate these performance benefits by modifying two well-known storage systems (Cassandra~\cite{la10cassandra} and ZooKeeper~\cite{hu10zookeeper}).
We plug \name\ on top of these, build three applications (a Twissandra-based microblogging service~\cite{twissandra}, an ad serving system, and a ticket selling system), and experiment on Amazon EC2.

Our evaluation first demonstrates that there is a sizable time window between preliminary and final views, which applications can use for speculation.
Second, using YCSB workloads A, B, and C, we show that we can reduce the latency of strongly consistent operations by up to $40\%$ (from $100ms$ to $60ms$) at little cost ($10$\% bandwidth increase, $6$\% throughput drop) in the ad system.
The other two applications exhibit similar improvements.
Even if the preliminary result is often inconsistent (25\% of accesses), incremental consistency incurs a bandwidth overhead of only 27\%.

In the rest of this paper, we overview our solution in the context of related work (\Cref{sec:overview}) and present the \name\ interface (\Cref{sec:abstraction}).
We show how applications use \name\ (\Cref{sec:action}), and describe the bindings to various storage stacks (\Cref{sec:bindings}).
We then give a comprehensive evaluation (\Cref{sec:evaluation}) and conclude (\Cref{sec:conclusions}).

\section{Overview \& Related Work}
\label{sec:overview}
%% In this paper, we address the issue of programming with replicated data.
%% We approach this problem with a novel data abstraction, \interface s.
This paper addresses the issue of programming and speculating with replicated objects through a novel abstraction called \name.
In this section, we overview the main concepts behind \interface s, and we contrast our approach with related work.

\subsection{Consistency Choices}
There is an abundance of work on consistency models.
These range from strong consistency protocols~\cite{lam98paxos,junq11zab,van04chain}, some optimized for WAN or a specific environment~\cite{Corbett:2013:Spanner,drago15farm,kras13mdcc,le15rifl,xie14salt,zh15tapir}, through intermediary models such as causal consistency~\cite{du2014gentlerain,Lloyd2013stronger}, to weak consistency~\cite{decandia2007dynamo,Terry1995Bayou}.
As a recent development, storage systems offer multiple---i.e., \emph{differentiated}---consistency guarantees~\cite{cooper2008pnuts,kra09ration,per15tunable}.
This allows applications in the above-mentioned gray zone to balance consistency and performance on a per-operation basis: the choice of guarantees depends on how sensitive the corresponding operation is.

Differentiated guarantees can take the form of SLAs~\cite{terry13pileus}, policies attached to data~\cite{kra09ration}, dynamic quorum selection for quorum-based storage systems such as Dynamo~\cite{decandia2007dynamo} or others~\cite{riak,la10cassandra}, or ad-hoc operation invariants~\cite{bail15feral}.
In practice%
%% \needsrev{, to simplify application design}
, two consistency levels often suffice: weak and strong~\cite{simpleDB,appengine}.
Sensitive operations (e.g., account creation or password checking) use the strong level, while less critical operations (e.g., remove from basket) use weak guarantees~\cite{kra09ration,terry13pileus,yu00design} to achieve performance.

For instance, in Gemini~\cite{li2012making}, operations are either Blue (fast, weakly consistent) or Red (slower, strongly consistent).
For sensitive data such as passwords, Facebook uses a separate linearizable sub-system~\cite{lu2015existential}.
Likewise, Twitter employs strong consistency for ``certain sets of operations''~\cite{manhattan}, and Google's Megastore exposes strong guarantees alongside read operations with ``inconsistent'' semantics~\cite{ba11megastore}.
Another frequent form of differentiated guarantees appears when applications bypass caches to ensure correctness for some operations~\cite{ajouxchallenges,nis13memcache}.

Given this great variety of differentiated guarantees, we surmise that applications can benefit from mixing consistency models.
The notable downside of this approach is that application complexity increases~\cite{li14automa}.
Developers must orchestrate different storage APIs and consider the interactions between these protocols~\cite{ajouxchallenges,bail15feral,wada11consumer}.
Our work subsumes results in this area.
We propose to hide different schemes for managing consistency under a common interface, \name , which can abstract over a varying combination of storage tiers and reduce application complexity.
In addition, we introduce the notion of \emph{incremental consistency guarantees} (\incguarantees ), i.e., progressive refinement of the result of a \emph{single} \mbox{operation}.

% , which we discuss next.

 % many modern applications rely on weakly-consistent storage services~\cite{decandia2007dynamo,manhattan,ajouxchallenges}.
% Some operations in these applications, however, are sensitive, requiring stronger consistency.
% Examples include account creation, password checking, or shopping cart checkout.
% This situation motivated the introduction of differentiated consistency guarantees (in the spirit of~\cite{mes11dss}) in replicated storage services, allowing applications to fine-tune each operation~\cite{kra09ration,terry13pileus,yu00design}.

% With this technique, applications can configure $R$ and $W$, which specify how many replicas to involve in a read or write operation, respectively, out of $N$ total replicas.
% To improve latency, it is common to use $R + W \le N$, providing weak consistency~\cite{bailis2012PBS}.
% With $R + W > N$, the system trades latency for better freshness.
% Thus, applications can ajust $R$ and $W$ to differentiate among guarantees.

\begin{table*}[t]
    \centering
    \small
    \renewcommand{\arraystretch}{0.9}
    \begin{tabular*}{\textwidth}{>{\centering}m{0.11\textwidth} m{0.35\textwidth} m{0.48\textwidth}}
    \hline
    Category & Synopsis & Applications and use cases \\
    \hline
    Weak\\ Consistency & Use the \textbf{weakest, but fastest} consistency model, e.g., by using partial quorums, or going to the closest replica or cache. No benefit from \incguarantees. & Computation on static (BLOBs) content, e.g., thumbnail generator for images and videos, accessing cold data, fraud analysis, disconnected operations in mobile applications, etc.\\
    \hline
    Strong Consistency & Use the \textbf{strongest} available model, e.g., by going to the primary replica. Applications require correct results. & Infrastructure services (e.g., load-balancing, session stores, configuration and membership management services), stock tickers, trading applications, etc. \\
    \hline
    Incremental Consistency Guarantees (ICG) & Use multiple, \textbf{incremental} models. Applications benefit from weakly consistent values (e.g., by speculating or exposing them), but prefer correct results. & E-mail, calendar, social network timeline, grocery list, flight search aggregation, online shopping, news reading, browsing, backup, collaborative editing, authentication and authorization, advertising, etc.\\
    \hline
    \end{tabular*}
    % \vspace{-.3cm}
    \caption{Different patterns and their corresponding use cases. Many applications can benefit from \incguarantees .}
    \label{tab:patterns}
    % \vspace{-.3cm}
\end{table*}

\subsection{\incguarantees: Incremental Consistency Guarantees}

%% \pending{
%% Empirical and theoretical studies reveal that weak consistency rarely exposes anomalies and provides strong guarantees with high probability~\cite{bailis2012PBS,lu2015existential}.
%% \needsrev{This evidence suggests that strong consistency should pose a \emph{small overhead} (compared to weaker models) to be tenable in practice.}
%% Building on this observation, our \name\ primitive supplies the result of an operation in multiple steps, using incremental consistency guarantees (\incguarantees).
%% This strategy allows applications to exploit the individual advantages of each consistency model.
%% Namely: the low-latency of weak consistency (to speculate on the output) and the correctness of strong consistency (to verify speculation).
%% Speculation hence allows to reap more performance from---and lower the overhead of---strong consistency.
%% }

Applications which use strong consistency---either exclusively or for a few operations---do so to avoid anomalous behavior which is latent in weaker models.
Interestingly, recent work reveals that this anomalous behavior is rare in practice~\cite{bailis2012PBS,lu2015existential}.
%Applications which demand correctness, however, cannot embrace this empirical fact without potentially exposing anomalies.
There are applications, however, which cannot afford to expose even those rare anomalies.

%% For instance, consider a storage system which has $1\%$ chance of exposing an inconsistent value, e.g., when accesing user passwords with weak consistency.
%% Then such a system could execute $99\%$ of accesses under weak guarantees, while the anomalous $1\%$ would run with strong consistency, yielding an overall correct behavior.
%% Yet in practice, if such a system demands correctness---as it should---then it is forced to access passwords with strong consistency \emph{all the time}.
%% We propose the \incguarantees\ strategy to help applications avert this dilemma, and pay for correctness only when anomalies actually occur.

For instance, consider a system storing user passwords, and say it has $1\%$ chance of exposing an inconsistent password.
%% Then such a system could execute $99\%$ of accesses under weak guarantees, while the anomalous $1\%$ would run with strong consistency, yielding an overall correct behavior.
If such a system demands correctness---as it should---then it is forced to pay the price for strong consistency on \emph{every} access, even though this is not necessary in 99\% of cases.
We propose \incguarantees\ to help applications avert this dilemma, and pay for correctness only when inconsistencies actually occur.

With \incguarantees , an application can obtain both weakly consistent (called \emph{preliminary}) and strongly consistent (called \emph{final}) results of an operation, one by one, as these become available.
While waiting for the final result, the application can speculatively perform further processing based on the preliminary---which is correct on expectation.
Following our earlier example, this would help hide the latency of strong consistency for $99\%$ of accesses.

% Speculation is a well-known optimization, with applications to other areas such as state machine replication~\cite{we09toler}.
% (a similar approach is applied to state machine replication by Wester et al. \cite{we09toler}).

The full latency of strong consistency is only exposed in case of misspeculation, when the preliminary and final values diverge because the preliminary returned inconsistent data~\cite{we09toler}.
These are the $1\%$ cases where strong consistency is needed anyway.
Speculation through \incguarantees\ can lessen the most prominent argument against strong consistency, namely its performance penalty.
With \incguarantees\ we pay the latency cost of strong consistency only when necessary, regardless of how often this is the case.

Speculation is a well-known technique for improving performance.
Traditionally, the effects of speculation in a system remain hidden from higher-level applications until the speculation confirms, since the effects can lead to irrevocable actions in the applications~\cite{kaprit12eve,mick10crom,night05speculator,we09toler}.
Alternatively, it has been shown that leaking speculative effects to higher layers can be beneficial, especially in user-facing applications, where the effects can be undone or the application can compensate in case of misspeculation~\cite{hel09quicksand,lang08experiences,lee15outatime,pa14planet}.
We propose to use eventual consistency as a basis for doing speculative work, as a novel approach for improving performance in replicated systems.
Also, more generally, we allow the application itself (which knows best), to decide on the speculation boundary~\cite{wes11operating}---whether to externalize effects of speculation, and later to undo or compensate these effects, or whether to isolate users from speculative state.

% In fact, even if less than $1\%$ of accessed objects are inconsistent, these are typically the most popular (``linchpin''~\cite{ajouxchallenges,nis13memcache}) objects, being both read- and write-intensive.
% Such anomalies have a disproportionate effect at application-level, since they reflect in many more than $1\%$ requests.

% Indeed, YCSB incorporates a \emph{Latest} request distribution to model popular items, where read activity is skewed towards recently updated items~\cite{co10ycsb}.
% Some systems are thus reluctant to give up strong consistency~\cite{Corbett:2013:Spanner}.

% , given that strong consistency is often sacrificed due to its high latency cost (possibly differing by orders of magnitude \cite{terry13pileus}).

% For applications in the gray zone between an easy choice of strong consistency (trading applications, membership management services, etc.) and an easy choice of weak consistency (computation on static/cold data, disconnected operations in mobile applications, etc.), the choice of a consistency level is hard.

% Even under these dire conditions, speculation is still efficient in terms of client bandwidth and computation overhead.

% With incremental consistency guarantees provided by \name,

% With \incguarantees, all operations resolve as fast as $R=1$, but whenever the weakly-consistent response lets stale data slip, the second (correct) response remediates it.
% In this sense, the multiple responses in \incguarantees\ complement each other.
% This strategy is essentially a form of speculation.

Besides speculation, \incguarantees\ is useful in other cases as well.
For instance, applications can choose dynamically whether to settle with a preliminary value and forsake the final value altogether.
This is a way to obtain application-specific optimizations, e.g., to enforce tight latency SLAs.
Alternatively, we can \emph{expose} the preliminary response to users and revise it later when the final response arrives. This strategy is akin to compensating in case of misspeculation, as mentioned earlier.

% This strategy is related to the guesses and apologies model of Helland and Campbell~\cite{hel09quicksand}: exposing the preliminary is akin to making guesses; an apology means amending the output.
% This approach found success in the PLANET model for transaction processing to deal with WAN unpredictability~\cite{pa14planet}.

Clearly, not all applications are amenable to exploiting \incguarantees .
In~\Cref{tab:patterns} we give a high-level account on three categories of applications: (1) those which have no additional benefit from strong consistency or \incguarantees ; (2) those which require correct results but are not amenable to speculation; and at last (3) applications that can obtain performance without sacrificing correctness by leveraging \incguarantees .

% \footnote{\emph{Adi:} Remember to cross-check this table with other parts of the paper. \textbf{Should we be apologetic \& make it clear that this is not a comprehensive study?}}

\subsection{Client-side Handling of \incguarantees }

To program with \incguarantees , applications need to wait asynchronously for multiple replies to an operation (where each reply encapsulates a different guarantee on the result) while doing useful work, i.e., speculate.
To the best of our knowledge, no abstraction fulfills these criteria.
To minimize the effort of programming with \incguarantees , we draw inspiration from \emph{Promises}, seminal work on asynchronous remote procedure calls in distributed systems~\cite{lisk88promises}.

A Promise is a placeholder for a value that will become available asynchronously in the future.
Given the urgency to handle intricate parallelism and augmenting complexity in applications, it is not surprising that Promises are becoming standard in many languages~\cite{guava,instagram,folly,er13your}.
We extend the binary interface of Promises (a value either present or absent) to obtain a multi-level abstraction, which incrementally builds up to a final, correct result.

The Observable interface from reactive programming can be seen as a similar generalization of Promises.
Observables abstract over asynchronous data streams of arbitrary type and size~\cite{me12rx}.
Our goal with \interface s, in contrast, is to grant developers access to consistency guarantees on replicated objects in a simple manner.
The ProgressivePromise interface in Netty~\cite{netty4promises} also generalizes Promises.
While it can indicate progress of an operation, a ProgressivePromise does not expose preliminary results.

%%%
%%%
%%% from related work
%%%
%%%

% \pending{\textbf{This doesn't really fit anywhere.}\\
% Another thriving subject is mitigating the complexity of programming with weak consistency.
% CRDTs, for instance, are weakly-consistent data types with the interesting property that all replicas converge monotonically to a common state, which avoid some anomalies of weak consistency, such as resurfaced deletions~\cite{sh11crdt}.
% CRDTs found adoption both in programming models, e.g. Lasp~\cite{meik15lasp}, and storage systems~\cite{zaw15swift}.
% Some further notable efforts go towards helping developers to reason about optimal consistency choices.
% These involve tools for automating and verifying consistency protocols, including proof rules~\cite{got16cause}, \textsc{sieve}~\cite{li14automa}, or Chapar~\cite{les16chapar} and high-level programming models such as \textsc{Quelea}~\cite{siv15declarative} or \texttt{BFTSim}~\cite{sin08bft}.
% Our work complements these results by attempting to simplify programmer efforts through a clean, consistency-based interface, and the \interface\ abstraction which reduces the complexity of working with replicated objects.
% }

% Line of work

%!TEX root = ../bundle.tex

\begin{figure}[t]
\vspace{-0.4cm}
\centerline{\includegraphics[width=\columnwidth]{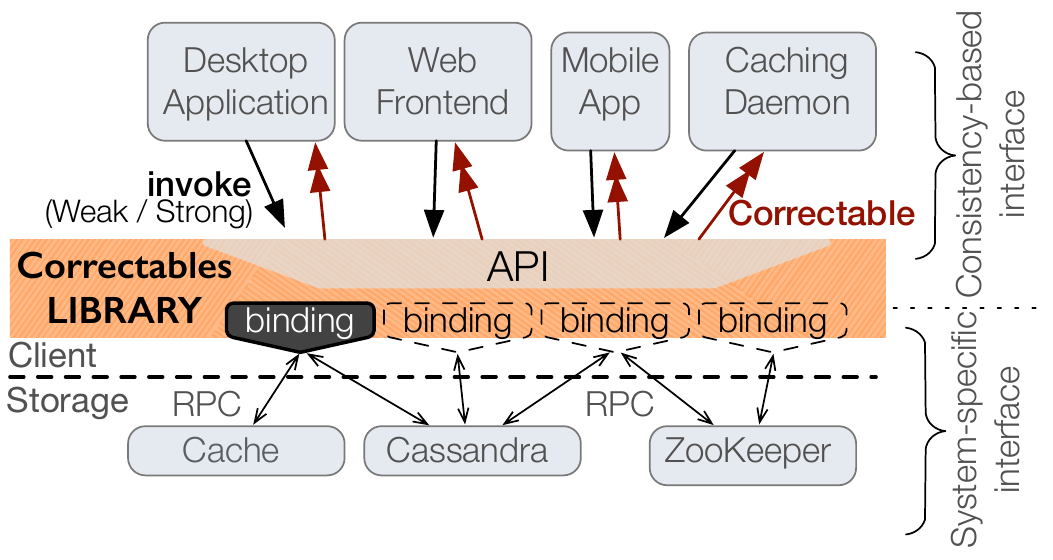}}
\caption{High-level view of \name , as an interface to the underlying storage.}
\label{fig:library}
\vspace{-0.4cm}
\end{figure}

\section{\name}
\label{sec:abstraction}

This section presents the \name\ interface for programming and speculating with replicated data.
Applications use this interface as a library, as~\Cref{fig:library} depicts.
At the top of this library sits the application-facing API.
The library is connected to the storage stack using a storage binding, which is a module that encapsulates all storage system specific interfaces and protocols.
\name\ fulfill two critical functions:
(i) translate API calls into storage-specific requests via a binding, and
(ii) orchestrate responses from the binding and deliver them---in an incremental way---to the application, using \emph{\interface} objects.
Each call to an API method returns a \interface\ which represents the progressively improving result (i.e. a result with \incguarantees).

% We start with a view on the high-level architecture, and we show how our library decouples applications from the underlying storage by adopting a thin, general consistency-based interface~(\Cref{sec:separating}).
% We then discuss \interface s, our solution to capture incremental consistency levels, and we also discuss a few relevant use-cases, showing how \name\ simplifies programming with replicated data~(\Cref{sec:correctables-interface}).
% Finally, we tie everything together by explaining the role of storage bindings in \name~(\Cref{sec:bindings}).

% \name\ stands between the application developer and the underlying stack, exposing to the developer an interface based on these levels.
% It is the job of library \emph{bindings} to implement concrete system-specific access methods which implement various consistency levels.

\begin{figure}[b]
\vspace{-.4cm}
\centerline{\includegraphics[width=0.85\columnwidth]{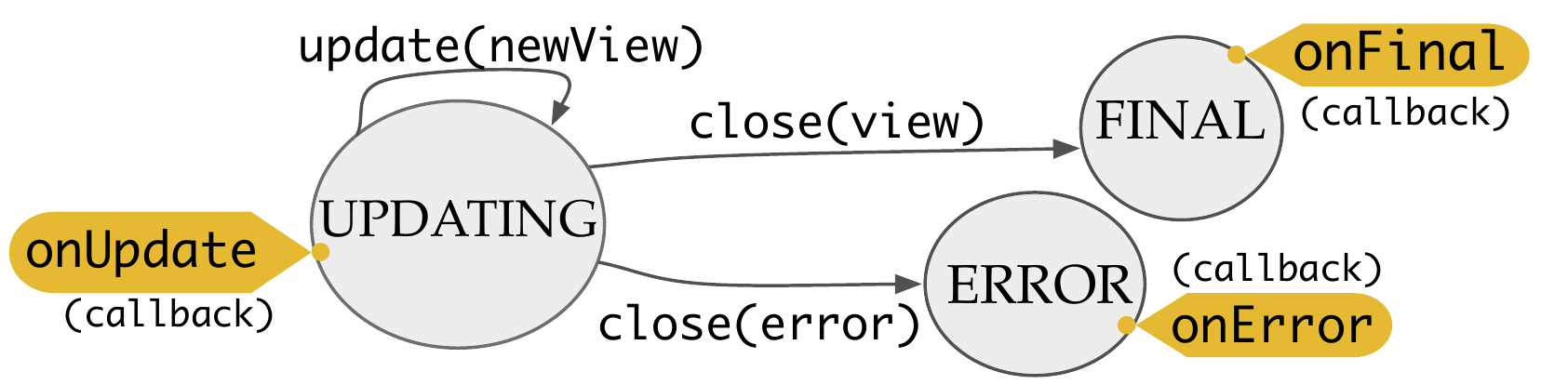}}
\caption{The three states, transitions, and callbacks associated with a \interface.}
\label{fig:states}
\end{figure}

\subsection{From Promises to \interface s}
\label{sec:correctables-interface}

As mentioned earlier, \name\ descend from Promises.
% \needsrev{A promise wraps asynchronous tasks---such as operations on replicated objects---and accepts callbacks which trigger upon task completion or error.}
To model an asynchronous task, a Promise starts in the \emph{blocked} state and transitions to \emph{ready} when the task completes, triggering any callback associated with this state~\cite{lisk88promises}.
Promises help with asynchrony, but not incrementality.
To convey incrementality, a \interface\ starts in the \emph{updating} state, where it remains until the final result becomes available or an error occurs (see \Cref{fig:states}).
When this happens, the \interface\ \emph{closes} with that result (or error), transitioning to the \emph{final} (or \emph{error}) state.
Upon each state transition, the corresponding callback triggers.
Preliminary results trigger a same-state transition (from \emph{updating} to \emph{updating}).
A \interface\ can have callbacks associated with each of its three states.
To attach these callbacks, we provide the \texttt{setCallbacks} method; together with \texttt{speculate}, these two form the two central methods of a \interface , which we examine more closely in~\Cref{sec:action}.

\subsection{Decoupling Semantics from Implementation}
\label{sec:access-API}

The \name\ abstraction decouples applications from storage specifics by adopting a thin, consistency-based interface, centered around \emph{consistency levels}.
This enables developers---who naturally reason in terms of consistency rather than protocol specifics---to obtain simple and portable implementations.
With \name, applications can transparently switch storage stacks, as long as these stacks support compatible consistency models.

%% This separation has two significant benefits.
%% First, developers remain agnostic to any technical details of the underlying storage system (how many replicas to access, which of them, coherence, etc.).
%% This is highly desirable, since developers reason about obtained results in terms of consistency guarantees.
%% Any requisite to translate guarantees into protocol specifics introduces additional complexity and sources of errors.
%% Second, application code remains the same---i.e., is portable---across different storage stacks, so we can switch transparently among them, as long as the consistency guarantees in these systems are interchangeable.
% The concrete choice of methods in our API reflects this decoupling:

Our API consists of three methods:
\begin{enumerate}
    \item \getfirst (\emph{operation}),
    \item \getlast (\emph{operation}), and
    \item \getinc (\emph{operation}[, \emph{levels}]).
\end{enumerate}

The first two allow developers to select either weak or strong consistency for a given \emph{operation}.
The returned \interface\ never transitions from \emph{updating} to \emph{updating} state and only closes with a final value (or error).
These two methods follow the traditional practice of providing a single result which lies at one extreme of the consistency/performance trade-off.

The third method provides \incguarantees , allowing developers to operate on this trade-off at run-time, which makes it especially relevant for applications in the above-mentioned gray area.
Instead of a single result (as is the case with the two former methods), \getinc\ provides incremental updates on the operation result.
Optionally, \getinc\ accepts as argument the set of consistency levels which the result should---one after the other---satisfy.
If this argument is absent, \getinc\ provides all available levels.
This argument allows some optimizations: e.g., if an application only requires a subset of the available consistency levels, this parameter informs a binding to avoid using the extraneous levels; we omit further discussion of this argument due to space constraints.
The available consistency levels depend on the underlying storage system and binding, which we discuss in more detail in \Cref{sec:bindings}.

%% To deliver asynchronous updates on the result of this method, our library exposes \interface\ objects.

In the next section, we show how to program with \name\ through several representative use-cases.
%% This discussion is relative to the canonical Promises of Liskov and Shrira~\cite{lisk88promises}.
In code snippets we adopt a Python-inspired pseudocode for readability sake.
For brevity we leave aside error handling, timeouts, or other features inherited from modern Promises, such as aggregation or monadic-style chaining~\cite{folly,er13your,lisk88promises}.

\section{\name\ in Action}
\label{sec:action}

This section presents examples of how \name\ can be useful on two main fronts.
(1) Decoupling applications from their storage stacks by providing an abstraction based on consistency levels.
(2) Improving application performance by means of \incguarantees, e.g., via  speculation or exploiting application-specific semantics.
%% \needsrev{In many cases, it is beneficial to even display inconsistent or stale data to the user, amending the output as soon as a better version becomes available, as is the case in~\needsref.}

\subsection{Decoupling Applications from Storage}

We first discuss a simple case of decoupling, where we argue for the first two functions in our API, namely \getfirst\ and \getlast.
As discussed in~\Cref{sec:overview}, many applications differentiate between weak and strong consistency to balance correctness with performance.
In practice, applications often resort to ad-hoc techniques such as cache-bypassing to achieve this, which complicates code and leads to errors~\cite{ajouxchallenges,er13your}.
\Cref{lst:redditIntro} shows code from Reddit \cite{reddit_usermsg}, a popular bulletin-board system and a prime example of such code.
Developers have to explicitly handle cache access (lines 6 and 9),
make choices based on presence of items in the cache (L7),
manually bypass the cache (L8) under specific conditions,
and write duplicate code (L12).

\begin{lstlisting}[mathescape,
                float,
                language=Python,
                label=lst:redditIntro,
                caption={Different consistency guarantees in Reddit~\cite{reddit_usermsg}, as an example of tight coupling between applications and storage. Developers must manually handle the cache and the backend.}]
from pylons import app_globals as g # cache access
from r2.lib.db import queries       # backend access

def user_messages(user, update = False):
  key = messages_key(user._id)
  trees = g.permacache.get(key)
  if not trees or update:
      trees = user_messages_nocache(user)
      g.permacache.set(key, trees) # cache coherence
  return trees
def user_messages_nocache(user):
  # Just like user_messages, but avoiding the cache...
\end{lstlisting}

\begin{lstlisting}[mathescape,
                language=Python,
                aboveskip=-5pt,
                belowskip=-6pt,
                float,
                floatplacement=H,
                caption={Reddit code rewritten using \interface s.},
                label=lst:reddit-correctable]
def user_messages(user, strong = False):
  key = messages_key(user._id)
  # coherence handled by invoke* functions in bindings
  if strong: return invokeStrong(get(key))
       else: return invokeWeak(get(key))
\end{lstlisting}

% \subsection{Using Differentiated Guarantees}
% In an alternative design, these functions could return a simple Promise representing the result.
% We choose \interface s, as those are first order citizens in our abstraction and, being a strict extension of Promises, can always be used as such.%

%% In an alternative design, these functions could block and return the result directly.
%% We find, however, that wrapping the result in a \interface\ is a more sensible choice.
%% The rationale behind this decision is that developers often need to tackle asynchrony, and \interface s are the ideal mechanism for this.
%% Since they descend from Promises, they inherit the benefits.
%% Notably: our API functions return a \interface\ and are thus non-blocking (an essential tool for scalability~\cite{joeerl03thesis,lisk88promises}),
%% and \interface s are first-order citizens---they may serve as arguments to functions or chained to form a pipeline~\cite{er13your}.%

%% Our JavaScript implementation, however, leverages the API and functionality of Promises/Futures in modern languages~\cite{guava,instagram,folly,er13your}.}
%% Using \interface s thus has significant advantages.

Instead of explicit cache-bypassing, we can employ \getfirst\ and \getlast\ to substantially simplify the code by replacing ad-hoc abstractions like \texttt{user\_messages} and \texttt{user\_messages\_nocache}, as \Cref{lst:reddit-correctable} shows.
Furthermore, we can replace other near-identical functions for differentiated guarantees, eliminating duplicate logic.%
\footnote{Similar pairs of ad-hoc functions exist in Reddit for accessing other objects.
  Perhaps accidentally, these other functions contain comments referring to \texttt{user\_messages} instead of their specific objects.
  We interpret this as a strong indication of ``copy-pasting'' code, which \name\ would help prevent.}
Cache-coherence and bypassing is completely handled by the storage-specific binding, reducing programmer effort and application-level complexity.

The third method in our library is \getinc .
\interface s are crucial for this method, since it captures \incguarantees.
\getinc\ allows applications to speculate on preliminary values (hiding the latency of strong consistency), or exploit application-specific semantics, as we show next.

\subsection{Speculating with \interface s}

Many applications are amenable to speculating on preliminary values to reap performance benefits.
To understand how to achieve this, we consider any non-trivial operation in a distributed application which involves reading data from storage.
Using \getinc\ to access the storage, applications can perform speculation on the preliminary value.
If this preliminary value is confirmed by the final value, then speculation was correct, reducing overall latency~\cite{we09toler}.
Examples where speculation applies include password checking or thumbnail generation (as mentioned in~\cite{terry13pileus}), as well as operations for airline seat reservation~\cite{yu00design}, or web shopping~\cite{kra09ration}.

\begin{lstlisting}[mathescape,
                language=Python,
                float,
                floatplacement=t,
                belowskip=-.6cm,
                label=lst:speculation-correctables,
                caption={Generic speculation with \interface s. The square brackets indicate that \texttt{abortFunc} is optional.}]
invoke(read(...))
  .speculate(speculationFunc[, abortFunc])
  .setCallbacks(onFinal = (res) => deliver(res))
\end{lstlisting}

\Cref{lst:speculation-correctables} depicts how this is performed in practice.
Even though such speculation can be orchestrated directly by using the \texttt{onUpdate} and \texttt{onFinal} callbacks of a \interface\ object, we provide a convenience method called \texttt{speculate} that captures the speculation pattern (L2).
It takes a speculation function as an argument, applying it to every new view delivered by the underlying \interface\ if this view differs from the previous one.
The \texttt{speculate} method returns a new \interface\ object which closes with the return value of the user-provided speculation function.
If the final view matches a preliminary one (which is the common case), the new \interface\ can close immediately when the final view becomes available, confirming the speculation.
Otherwise, it closes only after the speculation function is (automatically) re-executed with correct input.
In the latter case, an optional abort function is executed, undoing potential side-effects of the preceding speculation.
%% \needsrev{The \texttt{speculate} method makes sure to always invoke this function on the correct view, possibly starting its execution earlier, if a preliminary view matches the final one.}
Next, we discuss an ad serving system as an example application that can benefit from such speculation.

\paragraph{Advertising System.}
\label{sec:advertisement}
% Ad systems have conflicting requirements, requiring both good latency and consistency.
Typically, ads are personalized to user interests.
These interests fluctuate frequently, and so ads change accordingly~\cite{kor10cft}.
Given their revenue-based nature, advertising systems have conflicting requirements, as they aim to reconcile consistency (freshness of ads) with performance (latency)~\cite{cooper2008pnuts,Corbett:2013:Spanner}.
We thus find that they correspond to our notion of gray area, and are a suitable speculation use-case.

\Cref{lst:speculation-ads} shows how we can use \incguarantees\ while fetching ads.
First, we obtain a list of \emph{references} to personalized ads using the \getinc\ method (L2).
This method returns both a preliminary view (with weak guarantees) and a final (fresh) view.
Using the references in the preliminary view, we fetch the actual ads content and media, and do any post-processing, such as localization or personalization (L3).
If the final view corresponds to the preliminary, then speculation was correct, and we can deliver (L4) the ads fast; otherwise, \texttt{getAds} re-executes on the final view, and we deliver the result later.
We use this application as our first experimental case-study (\Cref{sec:tickets-case-study}).

The pattern of fetching objects based on their references---which themselves need to be fetched first---is widespread.
It appears in many applications, such as reading the latest news, the most recent transactions, the latest updates in a social network, an inventory, the most pressing items in a to-do list or calendar, and so on.
In all these cases, the application needs to chase a pointer (reference) to the latest data, while weak consistency can reveal stale values, which is undesirable.
We avoid stale data by reading the references with \getinc , and we mask the latency of the final value by speculatively fetching objects based on the preliminary reference.
% Since this is a widespread pattern, we discuss next a concrete application.

% All these cases are amenable to speculation, masking the latency of the first fetch by speculatively fetching objects that depend on it.

% We also consider augmenting this system to have a cache, since ads can also be displayed while offline~\cite{meik15lasp}.

\begin{lstlisting}[mathescape,
                language=Python,
                float,
                floatplacement=t,
                belowskip=-.4cm,
                label=lst:speculation-ads,
                caption={Example of applying speculation in an advertising system to hide latency of strong consistency.}]
def fetchAdsByUserId(uid):
  invoke(getPersonalizedAdsRefs(uid))
    .speculate(getAds) # fetch & post-process ads
    .setCallbacks(onFinal = (ads) => deliver(ads))
\end{lstlisting}

\subsection{Exploiting Application Semantics}

Applications can exploit their specific semantics to leverage the preliminary and the final values of \getinc .
For instance, consider the web auction system mentioned by Kraska et al.~\cite{kra09ration}, where strong consistency is critical in the last moments of a bid, but is not particularly helpful in the days before the bid ends, when contention is very low and anomalous behavior is unlikely.
Another example is selling items from a predefined stock of such items.
If a preliminary response suggests that the stock is still big, it is safe to proceed with a purchase.
Otherwise, if the stock is almost empty, it would be better to wait for the final response.
This is the case, for instance, for a system selling tickets to an event, which we describe next.
 % in more detail.

%% We find another subtler---yet broader---use of incremental guarantees by exploiting application semantics.
%% This is a well-known technique for coping with weak consistency.
%% \pending{\textbf{Ideally, we need a high-level description of what we mean by exploiting app. semantics, along the lines of: }\\We can exploit application semantics when they allow operations to be associative and commutative, and hence the strict ordering of strong consistency is not necessary~\cite{hel09quicksand}.}

\paragraph{Selling Tickets for Events.}
\label{sec:ticketApp}
%% It was previously observed that selecting consistency levels at run-time can be beneficial, e.g, to guarantee latency SLAs~\cite{terry13pileus}.
%% This can also help exploit periods when correctness can be ensured without requiring strong consistency.

%% We consider a ticket selling application.
%% This example helps us both illustrate how to exploit consistency levels dynamically, and how \name\ apply to queues.
%% As with speculation, the goal is to extract performance without sacrificing correctness.

For this application system, we depart from the popular key-value data type.
First, as we want to avoid overselling, we need a stronger abstraction to serialize access to the ticket stock.
Simple read/write objects (without transactional support) are fundamentally insufficient \cite{Herlihy91WS}.
Second, we want to demonstrate the applicability of \incguarantees\ to other data types.
We thus model the ticket stock using a queue, which is a simple object, yet powerful enough to avoid overselling.

%% In this application, we depart from the popular key-value data type.
%% The rationale is three-fold: with simple key-value objects, we cannot serialize access to the ticket stock and avoid overselling%
%% \footnote{To do so, we would need transactions. We believe transactions are an interesting avenue for future work on \incguarantees .};
%% we want to keep the application simple; and we are interested in applying \incguarantees\ to other data types.
%% We thus model the ticket stock using a queue.

Event organizers enqueue tickets and retailers dequeue them.
This data type allows us to serialize access to the shared ticket stock~\cite{aguil07sinfonia,kra09ration}.
We assume, however, that tickets bear no specific ordering (i.e., there is no seating).
Clients are interested in purchasing \emph{some} ticket, and it is irrelevant which exact element of the queue is dequeued.
We can thus resort to weak consistency most of the time, and use strong consistency sparingly.
We consider a weakly consistent result of an operation to be the outcome of simulating that operation on the local state of a single replica (see \Cref{sec:serverside}).

% In order to avoid overselling, we model the tickets using a replicated queue.
% We note that overselling is unavoidable without transactions in a simple key/value store.

\Cref{lst:ticketSellingBuy} shows how we can selectively use strong consistency in this case, based on the estimated stock size.
For each purchase, retailers use \getinc\ with the dequeue operation.
This yields a quick preliminary response, by peeking at the queue tail on the closest replica of the queue.
If the preliminary value indicates that there are many tickets left (e.g., via a ticket sequence number, denoting the ticket's position in the queue), which is the common case, the purchase can succeed without synchronous coordination on dequeue, which completes in the background.
This reduces the latency of most purchase operations.
As the queue drains, e.g. below a predefined threshold of $20$ tickets, retailers start waiting for the final results, which gives atomic semantics on dequeuing, but incurs higher latency.
This system represents our second experimental case study (\Cref{sec:tickets-case-study}).

\begin{lstlisting}[mathescape,
                language=Python,
                float,
                label=lst:ticketSellingBuy,
                belowskip=-.6cm,
                caption={Dynamic selection of consistency guarantees in a ticket selling system. If there are many tickets in the stock, we can safely use weak consistency.}]
def purchaseTicket(eventID):
  done = false
  invoke(dequeue(eventID)).setCallbacks(
    onUpdate = (weakResult) =>
      if weakResult.ticketNr > THRESHOLD:
        done = true # many tickets left, so we can buy
        confirmPurchase()
    onFinal = (strongResult) =>
      if not done and strongResult is not null:
        confirmPurchase() # we managed to get a ticket
      else: display("Sold out. Sorry!"))
\end{lstlisting}

% A limited number of tickets for some event is being sold on a first come first served basis.
% With a simple key-value store, overselling is unavoidable in presence of contention.
% To avoid overselling, we use a distributed queue (instead of simple read/write objects) to model the stock of available tickets.

% Since coordination is necessary to provide atomic semantics of dequeuing, the final (atomic) response arrives later.
% We can exploit the fact that it is irrelevant which exact element of the queue is removed when buying a ticket (we assume that personalized in formation is added to the ticket afterwards).
% Hence, the only problematic case is when the queue is almost empty; the preliminary response contains an element, but the final one is empty due to concurrent dequeuing.
% This case can be handled by waiting for the final value before proceeding with the purchase, if the queue size is small (e.g. 20 tickets).

% We investigate the impact of this approach quantitatively in \Cref{sec:tickets-case-study}

\subsection{Exposing Data Incrementally}

In some cases, it is beneficial to expose even incorrect (stale) data to the user if this data arrives fast, and amend the output as more fresh data becomes available.
Indeed, a quick approximate result is sometimes better than an overdue reply~\cite{decandia2007dynamo,terry13pileus}.
Many applications update their output as better results are available.
A notable example is flight search aggregators~\cite{skyscanner}, or generally, applications which exhibit high responsiveness by leaking to the user intermediary views on an ongoing operation~\cite{lang08experiences,lee15outatime}, e.g., previews to a video or shipment tracking.
We can assist the development of this type of applications, as we describe next.

\paragraph{Smartphone News Reader.}
Consider a smartphone news reader application for a news service replicated with a primary-backup scheme~\cite{terry13pileus}.
Additionally, recently seen news items are stored in a local phone cache.
With \incguarantees\ provided by \name, the application can be oblivious to storage details.
It can use a single logical storage access to fetch the latest news items, as \Cref{lst:newsReader} shows.
The binding would translate this logical access to three actual requests:
one to the local cache, resolving almost immediately,
one to the closest backup replica, providing a fresher view,
and one to a more distant primary replica, taking the longest to return but providing the most up-to-date news stories.

\begin{lstlisting}[mathescape,
                language=Python,
                label=lst:newsReader,
                float,
                belowskip=-.7cm,
                caption={Progressive display of news items using \name . The \texttt{refreshDisplay} function triggers with every update on the news items.}]
invoke(getLatestNews()).setCallbacks(
    onUpdate = (items) => refreshDisplay(items))
\end{lstlisting}

\subsection{Discussion: Applicability of ICG}
\label{sec:applicability}

In a majority of use-cases, we observe that two views suffice.
\name , however, support arbitrarily many views.
Note that this does not add any complexity to the interface and can be useful, as the news reader application shows.

There are other examples of applications which can benefit from multiple views.
A notable use-case are blockchain-based applications (e.g., Bitcoin \cite{nakamotobitcoin}), where \name\ can track transaction confirmations as they accumulate and eventually the transaction becomes an irrevocable part of the blockchain, i.e., strongly-consistent.
This is a use-case we also implemented, but omit for space constraints.
In larger quorum systems (e.g., BFT), \name\ can represent the majority vote as it settles.
Search or recommenders, likewise, can benefit from exposing multiple intermediary results in subsequent updates.\footnote{We are grateful to our anonymous OSDI reviewers for this particularly constructive idea.}

Intuitively, multiple preliminary views are helpful for applications requiring live updates.
On the one hand, several preliminary values would make the application more interactive and offer users a finer sense of progress.
This is especially important when the final result has high latency (Bitcoin transactions take tens of minutes).
%On the other hand, there is a downside.
%Simply put, as the replicated system delivers more preliminary views for an operation, less operations can be sustained: overall throughput drops.
On the other hand, as the replicated system delivers more preliminary views for an operation, less operations can be sustained and overall throughput drops.
Thus, applications which build on ICG with multiple incremental views observe a trade-off between interactivity and throughput. This trade-off can be observed even when the system delivers only two views (\Cref{sec:cassandra-gaps-load}).

In order to be practical, the cost of generating and exploiting the preliminary values of ICG must not outweight their benefits.
The cost of generating ICG is captured in the trade-off we highlighted above; the cost of exploiting ICG is highly application-dependent.
If used for speculation, the utility of $2+$ views depends on how expensive it is to re-do the speculative work upon misspeculation.
%Exploiting ICG means speculating on the preliminary values, so the utility of $2+$ views depends on how expensive it is to re-do the speculative work for evey \texttt{onUpdate}.
This can range from negligible (simply display preliminary views) to potentially very expensive (prefetch bulky data).
Additionally, the utility also depends on how often misspeculation actually occurs.
This depends on the workload characteristics: workloads with higher write ratios elicit higher rates of inconsistencies, and thus more misspeculations (\Cref{sec:cassandra-gaps-load}--\textbf{Divergence}).

% For doing speculation, two views are necessary and often ideal: the first provides a predictably good starting point for speculative work, and the second confirms of aborts that work.

There are also cases when using ICG is not an option.
This is either due to the underlying storage providing a unique consistency model and lacking caches, or due to application semantics, which can render ICG unnecessary---we give examples of this in the first two rows of~\Cref{tab:patterns}.
\name , however, are beneficial beyond ICG.
This abstraction can hide the complexity of dealing with storage-specific protocols, e.g., quorum-size selection.
The application code thus becomes portable across different storage systems.

\section{Bindings}
\label{sec:bindings}

Our library handles all the instrumentation around \interface\ objects.
This includes creation, state transitions, callbacks, and the API inherited from Promises~\cite{folly,er13your}.
%% Storage \emph{bindings} also bear a notable burden:
Bindings are storage-specific modules which the library uses to communicate with the storage.
These modules encapsulate everything that is storage system specific, and thus draw the separating line between consistency models---which \name\ expose---and implementations of these models. In this section, we describe the binding API, and show how bindings can facilitate efficient implementation of \incguarantees\ with server-side support.

\subsection{Binding API}

An instance of our library always uses one specific binding.
A binding establishes:
(1) the concrete configuration of the underlying storage stack (e.g., Memcache on top of Cassandra) together with
(2) the \emph{consistency levels} offered by this stack, and
(3) the implementation of any storage specific protocol (e.g., for coherence, choosing quorums). This allows the library to act as a client to the storage stack.

When an application calls an API method (\Cref{sec:access-API}), the library immediately returns a \interface .
In the background, we use the \emph{binding API} to access the underlying storage.
The binding forwards responses from the storage through an upcall to the library.
The library then updates (or closes) the associated \interface, executing the corresponding callback function.

The binding API exposes two methods to the library.
First, \texttt{consistencyLevels()} advertises to the library the supported consistency levels.
It simply returns a list of supported consistency levels, ordered from weakest to strongest.
In most implementations, this will probably be a one-liner returning a statically defined list.
The second function is \mbox{\texttt{submitOperation(op, consLevels, callback)}}.
The library uses this function to execute operation \texttt{op} on the underlying storage, with \texttt{consLevels} specifying the requested consistency levels.
The \texttt{callback} activates whenever a new view of the result is available.
The binding has to implement the protocol for executing \texttt{op} and invoke \texttt{callback} once for each requested consistency level.

% \texttt{consistencyLevels()}, used to advertise the supported consistency levels, and
% \texttt{submitOperation(operation, consLevels, callback)}, used to perform operations with various consistency levels on the storage system.
% \begin{compactitem}
% \item \texttt{consistencyLevels()} simply returns a list of supported consistency levels, ordered from weakest to strongest.
%   In most implementations, this will probably be a on-liner returning a statically defined list.
% \item \texttt{submitOperation(operation, consLevels, callback)} is used by \name\ to perform \texttt{operation} on underlying storage system.
%   The \texttt{consLevels} parameter is a set of consistency levels the result should have.
%   The callback is used to inform \name\ about the availability of results of the submitted operation.
%   The task of the binding is to perform \texttt{operation} on the actual storage system and call \texttt{callback} at least once for each consistency level in \texttt{consLevels}.
% \end{compactitem}

\Cref{lst:binding} shows the implementation of a simple binding for a primary-backup storage, supporting two consistency levels.
A more sophisticated binding could access the backup and primary in parallel, or could provide more than two consistency levels.
We designed the binding API to be as simple as possible; contributors or developers wishing to support a particular store must implement this API when adding new bindings.
We currently provide bindings to Cassandra and ZooKeeper.

%% \needsrev{We note that the \texttt{timeout} and \texttt{close} methods of \interface s require a way notify the backend of operation interruption.
%% Experience with Promises reveal that end-to-end interruption is non-trivial, but possible~\cite{er13your}.
%% We leave this for future work.}

%\footnote{The binding implementation in \Cref{lst:binding} is naive, since it does not access the backup and the primary in parallel.}

\begin{lstlisting}[mathescape,
                language=Python,
                float=b,
                label=lst:binding,
                caption={Simple binding to a storage system with primary-backup replication.}]
def consistencyLevels():
  return [WEAK, STRONG]

def submitOperation(operation, consLevels, callback):
  if WEAK in consLevels:
    backupResult = queryClosestBackup(operation)
    callback(backupResult, WEAK)
  if STRONG in consLevels:
    primaryResult = queryPrimary(operation)
    callback(primaryResult, STRONG)
\end{lstlisting}

\subsection{Efficiency and Server-side Support}
\label{sec:serverside}

On a first glance, \incguarantees\ might seem to evoke large bandwidth and computation overheads.
Indeed, if the \getinc\ method comprises multiple independent single-consistency requests, then storage servers will partly redo their own work.
Also, as the weakly and strongly consistent values often coincide, multiple responses are frequently redundant.
Such overheads would reduce the practicality of \incguarantees .

%% one uses \name\ directly on top of storage backends that only support simple differentiated consistency guarantees (a single consistency level per request), resources probably will be wasted.
%% Roughly, this happens on two fronts: (1) server-side request execution---redoing work that is the same for all consistency levels---and (2) client server communication---exchanging multiple lower-level requests and responses between client and server, substantial parts of which are expected to be identical [cite againg the empirical small divergence?] (especially if retrieving big objects).
%Minimizing server-side request execution overhead is rooted in the observation that

% depends on the way the coordination is performed.

With server-side support, however, we can minimize these overheads.
For instance, we can send a \emph{single} request to obtain all the incremental views on a replicated object.
An effective way to do this is to hook into the coordination mechanism of consistency protocols.
This mechanism is the core of such protocols, and the provided consistency model and latency depend on the type of coordination.
For example, asynchronous (off the critical path) coordination ensures eventually consistent results with low-latency~\cite{decandia2007dynamo}.
Coordination through an agreement protocol, as in Paxos~\cite{lam98paxos}, yields linearizability~\cite{herl90linearizability}, but with a higher latency.

\begin{figure}[b]
% \vspace{-0.4cm}
\centerline{\includegraphics[width=\columnwidth]{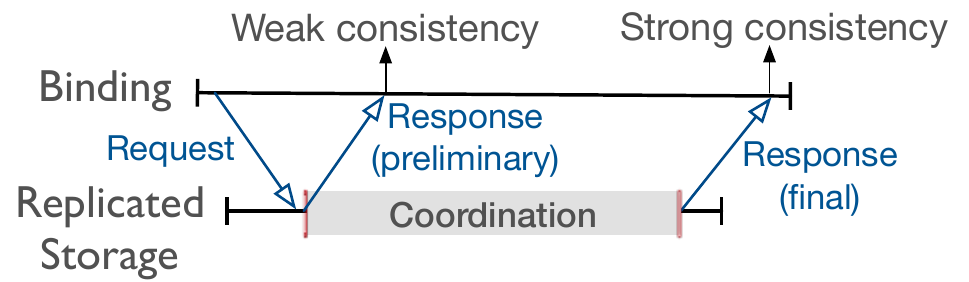}}
\caption{Simple server support for efficient \incguarantees . The storage system sends a preliminary response before coordinating. Note that for a single request, the storage provides two responses.}
\label{fig:serverSideEfficiency}
\end{figure}

Our basic insight is that we can get a good guess of the result already before coordinating, based on a replica's local state. In fact, this same state is being exposed when asynchronous coordination is employed, and as we alreay mentioned, this state is consistent on expectation.
The replica can leak a preliminary response---with weak guarantees---to the client prior to coordination (\Cref{fig:serverSideEfficiency}).
Moreover, we can reduce bandwidth overhead by skipping the final response if it is the same as the preliminary: a small \emph{confirmation} message suffices, to indicate that the preliminary response was correct.
Indeed, with such an optimization, \incguarantees\ has minor bandwidth overhead (\Cref{sec:cassandra-gaps}).

An additional benefit from this approach compared to sending two independent requests is that it prevents certain types of unexpected outcomes.
For instance, strong consistency might be more stale than weak consistency if responses to two independent requests were reordered by the WAN~\cite{terry13pileus}.
Using this approach, we modify two popular systems---Cassandra and ZooKeeper---to provide efficient support for \incguarantees.
Other techniques (e.g., master leases~\cite{cha07pml}) or replication schemes (e.g., primary-backup) can provide final views fast, skipping the preliminary altogether.

\paragraph{Cassandra.}
Cassandra uses a quorum-gathering protocol for coordination~\cite{gifford1979voting}.
In our modified version of Cassandra---called \ourcassandraname\ (\ourcassandra)---the coordinating node sends a preliminary view after obtaining the \emph{first} result from any replica.
This view has low latency, obtained either locally (if the coordinator is itself a replica) or from the closest replica.
Our binding to \ourcassandra\ supports two consistency levels, \emph{weak} (involving one replica) and \emph{strong} (involving two or more).
To minimize bandwidth overhead of \getinc, \ourcassandra\ uses the confirmation messages optimization we mentioned earlier.

\paragraph{ZooKeeper.}
To demonstrate the versatility of \name, we consider a different data type, namely replicated queues, which ZooKeeper can easily model~\cite{zkqueues}.
Our binding supports operations \texttt{enqueue} and \texttt{dequeue}, with weak and strong consistency semantics, accessible via \getfirst\ and \getlast, respectively.
The \getinc\ function supplies both consistency models incrementally.

The vanilla ZooKeeper implementation (\vanillazookeeper) has strong consistency~\cite{hu10zookeeper}.
For efficient \incguarantees, we implement \ourzookeepername\ (\ourzookeeper) by adding a fast path to \vanillazookeeper: a replica first simulates the operation on its local state, returning the preliminary (weak) result.
After coordination (via the Zab protocol~\cite{junq11zab}), this replica applies the operation and returns the strong response.
%% \needsrev{Under concurrent execution, it is possible that the weak and the strong results differ.}
% For example, concurrent \texttt{enqueue} is executed concurrently with other \emph{enqueue} requests, the weak and the strong results will differ.

% An \texttt{enqueue} operation, for example, returns the position of the new item in the queue.
%\getinc\ immediately returns the eventually-consistent response, which the contacted replica sends before coordinating with other replicas.
%% For atomicity (and the second response) this replica uses
%% the Zab protocol~\cite{junq11zab} to coordinates with the others, and returns the associated result.

\paragraph{Causal Consistency and Caching.}
We also implement a binding to abstract over a causally consistent store complemented by a client-side cache.
The \getinc\ function reveals two views: one from cache (very fast, possibly stale), and another from the causally consistent store.
This binding ensures write-through cache coherence, allows
cache-bypassing (\getlast ) or direct cache access (\getfirst ), e.g., in case of disconnected operations for mobile applications~\cite{per15tunable}.
Given the space constraints we focus on the two other bindings.

\section{Evaluation}
\label{sec:evaluation}

Our evaluation focuses on quantifying the benefits of \incguarantees.
Before diving into it, it is important to note that any potential benefit of \incguarantees\ is capped by performance gaps among consistency models.
Briefly, if strong consistency has the same performance as weaker models (or the difference is negligible) then applications can directly use the stronger model.
This is, however, rarely the case.
In practice, there can be sizable differences---up to orders of magnitude---across models~\cite{bailis2013highly,terry13pileus}.

We first describe our evaluation methodology, and then show that such optimization potential indeed exists.
We do so by looking at the performance gaps between weak and strong consistency in quorum-based (Cassandra) and consensus-based (ZooKeeper) systems.
We then quantify the performance gain of using \incguarantees\ in three case studies: a Twissandra-based microblogging service~\cite{twissandra}, an ad serving system, and a ticket selling application.

% two of them built on Cassandra which speculates on preliminary views, and a ticket selling system on top of a ZooKeeper queue which leverages application-specific semantics to benefit from \incguarantees.

% using microbenchmarks and YCSB, with two representative systems: Cassandra (quorum-based) and ZooKeeper (consensus-based).

%% We thus address the following questions:
%% (1) What is the exact potential for hiding latency through speculation on \incguarantees ?
%% (2) What are the actual benefits and costs of speculating through \incguarantees ?
%% We assess these concerns by studying two concrete case-studies---our first two applications (\Cref{sec:applications}).
%% We first discuss our experimental setup and then we proceed to these questions.

\subsection{Methodology}
\label{sec:experimental-setup}

%% We use a replication factor of $3$, with replicas geo-distributed in Frankfurt (FRK), Ireland (IRL), and N. Virginia (VRG).

\begin{figure}[b]
\centerline{\includegraphics[width=0.9\columnwidth]{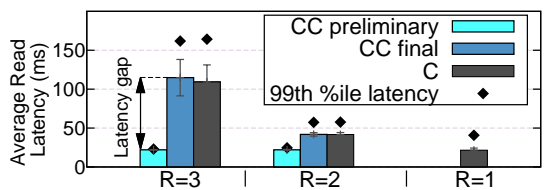}}
\caption{Single-request latencies in Cassandra for different quorum configurations. A bigger latency gap means a larger time window available for speculation.}
\label{fig:microbenchmark}
\end{figure}

% which service?
We run all experiments on Amazon's EC2 with m4.large instances and a replication factor of $3$, with replicas distributed in Frankfurt (FRK), Ireland (IRL), and N. Virginia (VRG).
%% Unless stated otherwise, we place the client in IRL, as other locations did not bring us substantially more insights.
Unless stated otherwise, to obtain WAN conditions, the client is in IRL and uses the replica in FRK; note that colocating the client with its contact server (i.e., both in IRL) would play to our advantage, as it would reduce the latency of preliminary responses and allow a bigger performance gap.
We also experiment with various other client locations in some experiments.

\begin{figure*}[t]
\centerline{\includegraphics[width=\textwidth]{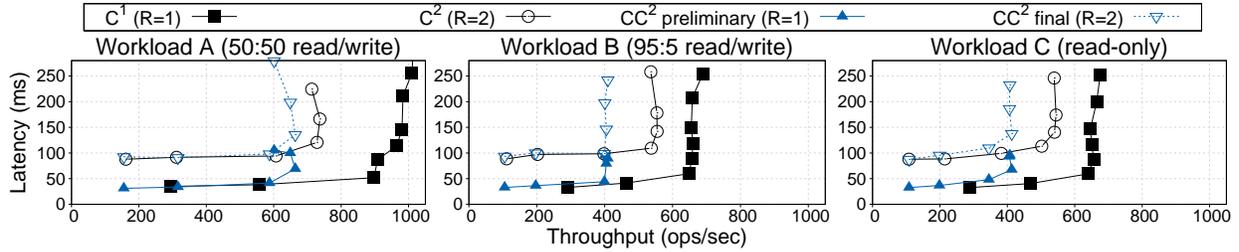}}
\caption{Performance of \ourcassandraname\ (\ourcassandra) compared to baseline Cassandra (\vanillacassandra). Note that the measurements for $CC^2$ have two results, one for the preliminary view and another for final. These two have the same throughput but different latencies.}
\label{fig:bobcat2}
\vspace{-.3cm}
\end{figure*}

For Cassandra experiments, we compare the baseline Cassandra v$2.1.10$ (labeled \vanillacassandra), with our modified \ourcassandraname\ (\ourcassandra).
We use superscript notation to indicate the specific quorum size for an execution, e.g., $C^1$ denotes a client reading from Cassandra with a read quorum $R=1$ (i.e., involving $1$ out of $3$ replicas).
For the ZooKeeper queue, we compare our modified \ourzookeepername\ (\ourzookeeper) against vanilla ZooKeeper (\vanillazookeeper), v$3.4.8$.
The cumulative implementation effort associated with \ourcassandra\ and \ourzookeeper , including three case studies, is modest, at roughly $3k$ lines of Java code.

% \pending{If not stated otherwise, standard deviations are negligible and we omit them for better readability.

% We demonstrate the latency gap between between the preliminary and the final results of an enqueue operation with various server configurations,
% For the ticket selling application, the we place a \ourzookeeper\ leader replica to IRL.
% \textbf{Move this somewhere else:}
% Clients, also located in IRL, connect to a \ourzookeeper\ follower in FRK to simulate WAN conditions.
% Each data point in a plot is an average over five trials.}

%% \paragraph{Systems.}
% \pending{To evaluate \name, we implement and experiment with bindings to three different storage systems: Cassandra, ZooKeeper, and \causalstore, our own causally consistent storage system.
% We create \ourcassandraname\ (\ourcassandra), an adaptaition of Cassandra with native \name\ support.
% We compare the use of \name\ with \ourcassandra\ to native access to vanilla Cassandra using microbenchmarks, YCSB workloads, and two applications: an ad serving system and a notification system backend.
% To explore the use of \name\ for other objects than simple key/value pairs, we implement a binding to ZooKeeper that exposes a queue interface, and evaluate an application for selling tickets on top of it.\\
% \textbf{Causally consistent social app?}
% }

\subsection{Potential for Exploiting \incguarantees }

To determine the potential of \incguarantees , we examine their behavior in practice.
% \needsrev{Previous studies reveal that there are gaps.}
%% Other studies of geo-replicated deployments have similar conclusions~\cite{bailis2013highly,terry13pileus}.
Studies show that large load on a system and high inter-replica latencies give rise to large performance gaps among consistency models~\cite{bailis2013highly,terry13pileus}.
To the best of our knowledge, however, there are no studies which consider a combination of incremental consistency models in a single operation.
We first investigate this behavior in Cassandra and then in ZooKeeper.

\subsubsection{Potential for Exploiting \incguarantees\ in Cassandra}
\label{sec:cassandra-gaps}

Cassandra can offer us insights into the basic behavior of \incguarantees\ in a quorum system.
As explained in ~\Cref{sec:bindings}, \ourcassandra\ offers two consistency models: weak,
which yields the \emph{preliminary} view ($R=1$), and strong, giving the \emph{final} view ($R=2$ or $R=3$, depending on the requested quorum size).
For write operations, we set $W=1$.
% \footnote{Experience from industry deployment of quorum-based systems reports that they ``did not observe staleness'' when using $R=2, W=1$~\cite{bailis2012PBS}.}
We use microbenchmarks and YCSB~\cite{co10ycsb} to measure single-request latency and performance under load, respectively.
For each \ourcassandra\ experiment, we run three 60-second trials and elide from the results the first and last $15$ seconds.
We report on the average and 99th \%ile latency, omitting error bars if negligible.
% For YCSB, we use the built-in measurements support.

\paragraph{Single-request Latency.}

% When using $CC^x$, a client expects two replies: a preliminary and a final view. The first should have the latency and consistency characteristics of $R=1$, and final views should evince the latency and consistency traits of $R=x$.
%, for a client in FRK connecting to the replica in IRL.!
We use a microbenchmark consisting of read-only operations on objects of $100B$.
We are interested in the performance gap between preliminary and final views as provided by \incguarantees , and we contrast these with their vanilla counterparts.
We thus compare $CC^2$ ($R\in\{1,2\}$) and $CC^3$ ($R\in\{1,3\}$) with $C^1$ ($R=1$), $C^2$ ($R=2$), and $C^3$ ($R=3$).
For $CC$, $R$ has two values: the read quorum size for the preliminary (weak) and for the final (strong) replies, respectively.

\Cref{fig:microbenchmark} shows the results for all these configurations, grouped by their read quorum size.
The average latency of preliminary views---whether it is for $CC^2$ or $CC^3$---follows closely the latency of $C^1$, which coincides with the $20ms$ RTT between client and coordinator replica.
Preliminary views reflect the local state on the replica in FRK, having the same consistency as $C^1$.
Final views of $CC^2$ and $CC^3$ follow the trend of the requested quorum size and reflect the behavior of $C^2$ and $C^3$ respectively.

The performance gap between the preliminary and final view for $CC^2$ is $20ms$.
The coordinator (FRK) is gathering a quorum of two: itself and the closest replica (IRL). The gap indeed corresponds to the RTT between these two regions.
For $CC^3$, the gap is much larger: up to $140ms$ for the 99th \%ile, due to the larger distance to reach the third replica (VRG).
By speculating on the preliminary views, applications can hide up to $20ms$ (or $140ms$) of the latency for stronger consistency.
In practice, such differences already impact revenue, as users are highly-sensitive to latency fluctuations~\cite{decandia2007dynamo,ham09latency}.

\paragraph{Performance Under Load.}
\label{sec:cassandra-gaps-load}
We also study the performance gap using YCSB workloads A (50:50 read/write ratio), B (95:5 read/write ratio), and C (read-only)~\cite{co10ycsb}.
To stress the systems and obtain WAN conditions, we deploy $3$ clients, one per region, with each client connecting to a remote replica. For brevity, we only report on the results for the client in IRL and $R=\{1,2\}$.
\Cref{fig:bobcat2} presents the average latency as a function of throughput.
We plot the evolution of both the preliminary and final views individually.
% We also show on the bottom row the 99th \%ile latency as we increase the load, as it helps us interpret the results.

We observe that \ourcassandra\ trades throughput due to the load generated on the coordinator, which handles \incguarantees .
We observe this behavior in all three workloads.
%% , and latency of final views suffers compared to the baseline.
This is to be expected, considering the modifications we did to implement preliminary replies (\Cref{sec:serverside}).
Briefly, we add another step to every read operation that uses quorums larger than one.
This step, called \emph{preliminary flushing}, occurs at any coordinator replica serving read operations as soon as that replica finishes reading the requested data from its local storage---and prior to gathering a quorum from other replicas.
% \needsrev{This step entails simply a flushing of the preliminary reply.}
This step generates additional load on the coordinator replica, explaining the throughput drop of $CC^2$ compared to baselines.
Work on replicated state machines (RSM) suggests an optimization~\cite{we09toler} which resembles our flushing technique. Perhaps unsurprisingly, the optimized RSM exhibits a similar throughput drop~\citep[\S6.2]{we09toler} as we notice in these experiments.

%% and higher latencies of final views
% , but not the latency improvement of preliminary views.

% We explain the latency decline of preliminary views by the observation that flushing is not on the critical path of reads for $R > 1$: replicas flush while waiting for a quorum.
% For $R = 1$, however, no quorum gathering happens and the system throughput is limited by the performance of a single replica (and the bandwidth of the client-server connection).
% When this throughput limit is reached, latency increases due to the replica being overloaded, i.e. the replica is the bottleneck.

% For $R > 1$, the throughput required to overload a replica is not reached, since the bottleneck is the inter-replica WAN bandwidth (which is irrelevant for $R = 1$).
% Thus, for $R > 1$, the replica has enough slack to flush preliminary views fast.
% We can still observe the bottleneck posed by the coordinating replica in the bottom row of \Cref{fig:bobcat2}, when the preliminary view latency starts increasing.
%% This behavior becomes more acute as the load increases over $60$ client threads and throughput saturates.

The latency gap between preliminary and final views is the same as the one we observe in the microbenchmarks.
To conclude, our results confirm that the performance gaps while using \incguarantees\ are noticeable, and hence there is room for hiding latency.

\paragraph{Divergence.}
\label{sec:divergence-benchmark}

To obtain more insight about the behavior of \incguarantees, we use \ourcassandra\ and the YCSB benchmark to measure how often preliminary values diverge from final results.
% We call this metric \emph{divergence}.
% : how often a weakly-consistent value (reading with $R=1$) is different than stronger consistency (reading with $R=2$) in a system of $3$ replicas.
We achieve this by using \getinc\ and comparing the preliminary view to the final one.
We run this experiment with a small dataset of $1K$ objects.
We aim at obtaining the conditions of a highly-loaded system where clients are mostly interested in a small (popular) part of the dataset.
% This should exacerbate contention and increase divergence.

\begin{figure}[b]
\vspace{-.3cm}
\centerline{\includegraphics[width=\columnwidth]{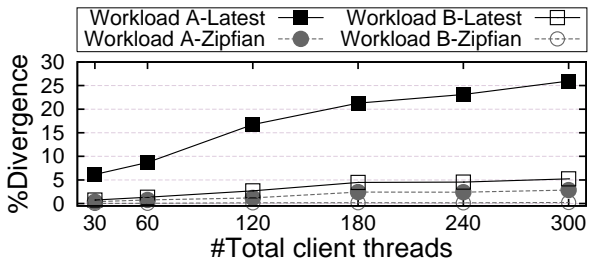}}
\caption{Divergence of preliminary from final (correct) views in Correctable Cassandra with various YCSB configurations.}
\label{fig:divergence}
\end{figure}

\Cref{fig:divergence} shows our result for a mix of representative YCSB workloads (A and B) and access patterns (Zipfian and Latest) with default settings.
% We find that some workloads can exhibit high divergence.
Notably, workload A (50:50 read/write) under Latest distribution (read activity skewed towards recently updated items) exhibits high divergence, up to $25\%$.
Under such conditions, using $R=1$ would yield many stale results.
Indeed, some applications with high write ratios, e.g., notification or session stores~\cite{co10ycsb,yam11riak}, tend to use $R=2$, even though this forces \emph{all} read operations to pay the latency price~\cite{bailis2012PBS}.

In fact, even if less than $1\%$ of accessed objects are inconsistent, these are typically the most popular (``linchpin''~\cite{ajouxchallenges,nis13memcache}) objects, being both read- and write-intensive.
Such anomalies have a disproportionate effect at application-level, since they reflect in many more than $1\%$ application-level operations.
Applications with high update ratios as modeled by workload A, e.g., social networks~\cite{cooper2008pnuts}, can thus benefit from exploiting \incguarantees\ to avoid anomalies.

\paragraph{Bandwidth Overhead.}

In addition to the throughput drop mentioned above, client-replica bandwidth is the next relevant metric which \incguarantees\ can impact.
Yet, optimizations can cut the cost of this feature (\Cref{sec:serverside}).
We implement such an optimization in \ourcassandra, whereby a final view contains only a small confirmation---instead of the full response---if it coincides with the preliminary view.
We note that in all experiments thus far we did not rely on this optimization, which makes our comparisons with Cassandra conservative.

\begin{figure}[t]
\centerline{\includegraphics[width=\columnwidth]{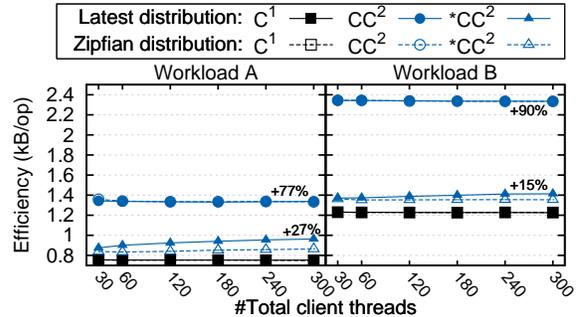}}
\caption{Efficiency (bandwidth overhead) of the \incguarantees\ implementation in Correctable Cassandra (CC).}
\label{fig:efficiency-combined}
\vspace{-.3cm}
\end{figure}

To obtain a worst-case characterization of the costs of \incguarantees , we consider the scenario where divergence can be maximal, as this will lessen the amount of bandwidth we can save with our optimization.
Hence, we consider the exact conditions we use in the divergence benchmark, where we discovered that divergence can rise up to $25\%$.
In this experiment, we measure the average data transferred (KB) per operation.
We contrast three scenarios.
First, as baseline, we use $C^1$, where clients request a single consistency version using weak reads.
The other two systems are $CC^2$ (without optimization) and $^*CC^2$ (optimized).

\Cref{fig:efficiency-combined} shows our results.
As expected, if divergence is very high---notably in workload A---then many preliminary results are incorrect.
% This means that we miss-speculate, and the size of finals cannot be reduced, increasing the data cost by up to $27\%$.
This means that final views cannot be replaced by confirmations, increasing the data cost by up to $27\%$.
Without any optimization, this would drive the cost up by $77\%$.
Workload B has a smaller write ratio ($5\%$), so a lower divergence and more optimization potential: we can reduce the overhead from $90\%$ down to $15\%$ (since most final views are confirmations).

%% We believe this proves that \incguarantees\ have a modest cost in terms of data usage.
Our experiments prove that \incguarantees\ have a modest cost in terms of data usage.
This cost can be further reduced through additional techniques (\Cref{sec:serverside}).
We remark that our choice of baseline, $C^1$, is conservative, because $CC^2$ offers better guarantees than $C^1$.
A different baseline would be a system where clients \emph{send two requests}---one for $R=1$ and one for $R=2$---and \emph{receive two replies}.
While such a baseline offers the same properties as $CC^2$, it would involve bigger data consumption, putting our system at an advantage.

\subsubsection{Potential for Exploiting \incguarantees\ in ZooKeeper}
\label{sec:zookeeper-gaps}

\begin{figure}[t]
\centerline{\includegraphics[width=\columnwidth]{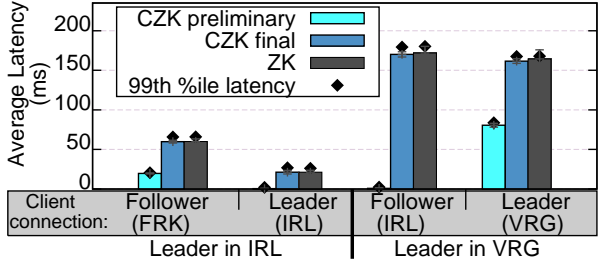}}
\caption{Latency gaps between preliminary and final views on the result of dequeue operations in Correctable ZooKeeper (\ourzookeeper) compared to ZooKeeper (\vanillazookeeper). Client is in IRL.}
\label{fig:zookeeper-enqueue}
\vspace{-.3cm}
\end{figure}

\paragraph{Latency Gaps.}
We also measure performance gaps in ZooKeeper queues for various locations of the leader and the replica which the client (in IRL) connects to.
We show the results for four representative configurations for adding elements to a queue (we discuss dequeuing in the context of a ticket selling system in~~\Cref{sec:tickets-case-study}).
The elements are small, containing an identifier of up to $20B$ (e.g., ticket number).
\Cref{fig:zookeeper-enqueue} shows the latency gaps when we use \incguarantees\ in \ourzookeepername\ (\ourzookeeper) compared to baseline ZooKeeper (\vanillazookeeper).

In all cases, the latency of the preliminary view (containing the name of the assigned znode) corresponds to the RTT between the client and the contacted replica.
This latency ranges from $2ms$ (when client and replica are both in IRL), through $20ms$ (the RTT from IRL to FRK), up to $83ms$ (the RTT between IRL and VRG).
The most appealing part of this result is perhaps the substantial gap which appears when the client and the closest follower are in IRL and the leader is distant (in VRG), in the third group of results in \Cref{fig:zookeeper-enqueue}.

\paragraph{Bandwidth Overhead.}
Storing big chunks of data is not ZooKeeper's main goal.
The client-server bandwidth is usually not dominated by the payload, reducing the benefits of the confirmation optimization.
For enqueuing, the bandwidth cost thus increases by roughly $50$\%, from $270$ to $400$ bytes/operation.
As expected, this corresponds to one additional (preliminary) response message in addition to the original request and (final) response.

\begin{figure}[t]
\centerline{\includegraphics[width=\columnwidth]{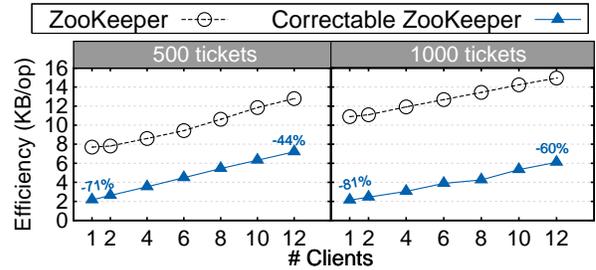}}
\caption{Efficiency (bandwidth overhead) for dequeuing operation in Correctable ZooKeeper (CZK) and ZooKeeper (ZK). Overhead in CZK is independent of queue size.}
\label{fig:tickets-efficiency}
\vspace{-.3cm}
\end{figure}

While queues are a common ZooKeeper use-case, a problem appears in standard dequeue implementations due to message size inflation~\cite{netflixqueue}.
Specifically, clients would first read the \emph{whole queue} and then remove the tail element.
To evade this problem in \ourzookeeper, clients only read the constant-sized tail relevant for dequeuing.
% \needsrev{This optimization is applicable to other cases besides \incguarantees\ or other data types, such as stacks. (What data types? The queue implementation is not directly transferable to a stack implementation...)}
\Cref{fig:tickets-efficiency} compares the bandwidth cost per dequeue operation in \ourzookeeper\ and \vanillazookeeper\ for different queue sizes as we increase the number of contending threads.
While the cost still increases with contention in both cases, in \ourzookeeper\ we make it independent of queue size, which is not the case for \vanillazookeeper.
As future work, we plan to make the dequeue cost also independent of contention using tombstones~\cite{sai05optimistic}.

%% Specifically, in ZooKeeper clients typically receive from the service the whole queue content (which is a ZNode) during an operation.
%% Given that queues may be reach thousands of items, this is not scalable.
%% Indeed, Netflix published their own version of a distributed queue on ZooKeeper, highlighting this very issue~\cite{netflixqueue}.

% Next, we quantify the usefulness of \incguarantees\ in two concrete applications.

\subsection{Case Studies for Exploiting \incguarantees}

Given the optimization potential explored so far, we now investigate how to exploit it in the context of three applications: the Twissandra microblogging service~\cite{twissandra}, an ad serving system, and a ticket selling system.
The first two build on \ourcassandra\ and use speculation. The last application uses \ourzookeeper\ queues.

%  We speculate using \incguarantees\

% and we speculate

%  and which we modify to use \incguarantees . (2) An ad serving system using \ourcassandra , and a ticket selling system on top of \ourzookeeper\ queues.
% We measure the \needsrev{actual} derived performance benefits at application level.
% We measure X and discus each case study in turn.
% and the costs in terms of bandwidth overhead (costs).

\subsubsection{Speculation Case Studies}
\label{sec:speculation}

For Twissandra, we are interested in \texttt{get\_timeline} operation, since this is a central operation and is amenable to optimization through speculation.
This operation proceeds in two-steps: (1) fetch the timeline (tweet IDs), and then (2) fetch each tweet by its ID.
We re-implement this function to use \getinc\ on step (1) and leverage the preliminary timeline view to speculatively execute step (2) by prefetching the tweets.
If the final timeline corresponds to the preliminary, then the prefetch was successful and we can reduce the total latency of the operation.
In case the final timeline view is different, we fetch the tweets again based on their IDs from this final view.

\begin{center}
\begin{figure*}[t]
\centerline{\includegraphics[width=\textwidth]{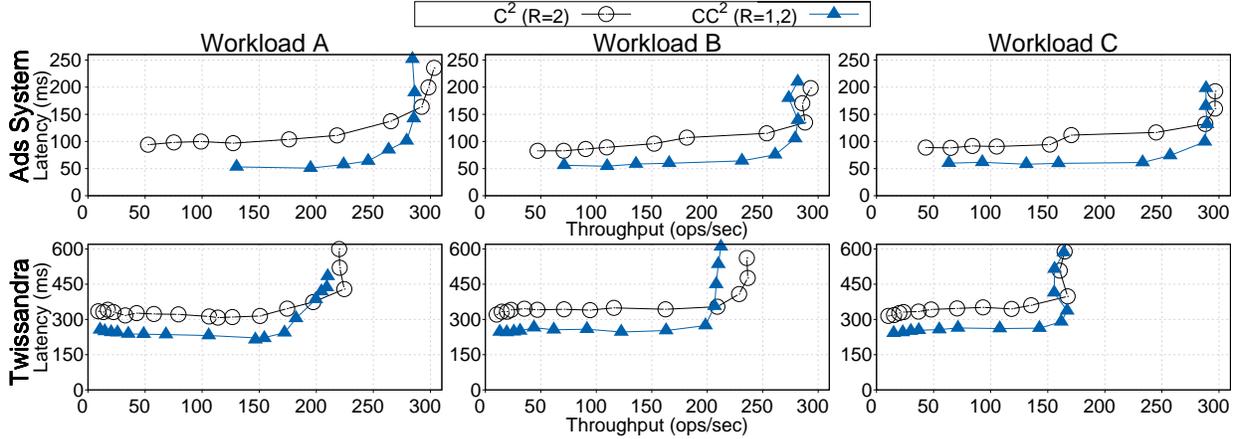}}
\caption{Using speculation via \incguarantees\ to improve latency in the advertising system and in Twissandra (\texttt{get\_timeline} operation). Correctable Cassandra (CC) improves latency by up to $40\%$ in exchange for a throughput drop of $6\%$.}
\label{fig:twissandra}
% \vspace{-.3cm}
\end{figure*}
\end{center}

\vspace{-1.2cm}

Our second speculation case study is the ad serving system we describe in~\Cref{sec:advertisement}.
The goal is to reduce the total latency of \texttt{fetchAdsByUserId} operation without sacrificing consistency, so we exploit \incguarantees\ by speculating on preliminary values (\Cref{lst:speculation-ads}).

For both systems, we adapt their respective operations to use \getinc\ ($R=\{1,2\}$) and plug them in the YCSB framework.
We compare these operations using a baseline that uses only the strongly consistent result ($R=2$), and does not leverage speculation.
For Twissandra we use a corpus of $65k$ tweets~\cite{tweetsdataset} spread over $22k$ user timelines; the ad serving system uses a dataset of $100k$ user-profiles and $230k$ ads, where each profile references between $1$ and $40$ random ads.

The results are in~\Cref{fig:twissandra}.
In contrast to our other experiments, we deploy Twissandra replicas in Virginia, N. California, and Oregon EC2 regions.
The goal is to see how performance gains vary based on deployment scenario.
The ads system uses the same configuration as before.
The client is in IRL for both experiments.

% As discussed earlier, ad systems have conflicting requirements, for both latency and consistency.

% \vspace{-1cm}

We first explain the results for the ads system.
As can be seen, these are consistent with our earlier findings from Cassandra experiments (\Cref{fig:bobcat2}).
We trade throughput for better latency.
Prior to saturation, we can serve ads with an average latency of $60ms$.
In the same conditions, the baseline achieves $100ms$ average latency (improvement by $40\%$).
In turn, the throughput drop is most noticeable in workload A, by $18ops/sec$ (reduced by $6\%$).
The smaller throughput drop compared to the raw results of \Cref{fig:bobcat2} is explained by the fact that each \texttt{fetchAdsByUserId} entails two storage accesses.
Only the first access, however, uses \incguarantees\ (to speculate).
The second storage access is hidden inside \texttt{getAds} (\Cref{lst:speculation-ads}, L3); this is a read with $R=2$, incurring no extra cost.

For Twissandra, we observe a lower throughput and higher latency, as the client is farther from coordinator and replicas are also more distant from each other.
But otherwise we draw similar conclusions.
Notably, across both of these case-studies, divergence was consistently under $1\%$, so the applications encountered very few misspeculations.

\begin{figure}[t]
% \vspace{-.4cm}
\centerline{\includegraphics[width=\columnwidth]{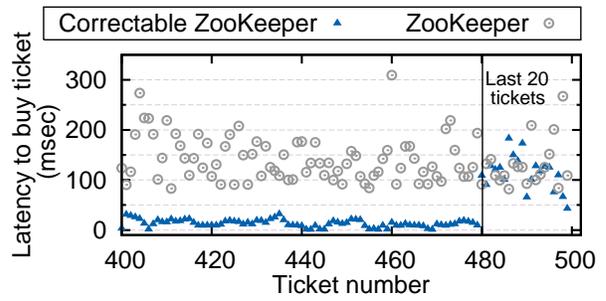}}
\caption{Selling tickets with \vanillazookeeper\ and \ourzookeeper. The last $20$ tickets incur high latency due to strong consistency.}
\label{fig:tickets-timeline}
\end{figure}

% \begin{center}
% \begin{figure*}[ht]
% \centerline{\includegraphics[width=0.9\textwidth]{./figures/news-perf}}
% \caption{Using speculation via \incguarantees\ to improve latency in an ads serving system. }
% \label{fig:news-perf}
% \end{figure*}
% \end{center}

\subsubsection{Selling Tickets to Events}
\label{sec:tickets-case-study}

A second notable use-case of \incguarantees\ is exploiting application semantics, as we discuss in the ticket selling system from~\Cref{sec:ticketApp} (see \Cref{lst:ticketSellingBuy}).
Here we exploit the fact that the position of a ticket in the queue is irrelevant.
%% ticket stock (queue) drains linearly.
Thus, in the common case, we can rely on the preliminary value.
Strong consistency (atomicity), however, becomes critical when ticket retailers are contending over the last few remaining tickets.
Using \incguarantees, we can switch dynamically between using the preliminary or the final results when the stock becomes low, to avoid overselling.
%% \Cref{lst:ticketSellingBuy} describes this functionality, which is the central point of our experiments.

% To quantify the performance gains, we experiment with the ticket-selling application on top of a \ourzookeeper\ queue, as described in .
%% For instance, in the ticketing application, we can already signal success to the user once the preliminary view resolves.
%% Even if the exact position of the ticket in the queue may change when the final view resolves, we can guarantee that the user will obtain a ticket.
We consider $4$ retailers concurrently serving (dequeuing) tickets from a fixed-size stock of $500$ tickets.
Retailers are colocated with a \ourzookeeper\ follower in FRK, the leader being in IRL.
We wait for the final (atomic, equivalent to \vanillazookeeper) response for the last $20$ tickets, otherwise we use the preliminary one.
This is a conservative bound; in our experiments, only the last two tickets were ``revoked'' by the final view on average, with a maximum of six.

\Cref{fig:tickets-timeline} shows individual ticket purchase latencies, averaged over five runs, compared to latencies with vanilla \ourzookeeper.
%% We reduce the overall average latency
% (including the last $20$ tickets)
%% from $142ms$ to $33ms$ by only paying for consistency for the last 20 tickets.
As long as there are more than $20$ tickets left, we reduce the purchase latency substantially.
The high variability of final view latencies is caused by contention between the retailers, which does not affect preliminary views.
We experiment also with larger ticket stocks ($1000$), but the queue length has no practical effect on latencies.
To support more contention (more retailers) in practice, such a ticketing service can scale-out.
For instance, we can shard the ticket stock and instantiate multiple replicated \ourzookeeper\ services, each of them serving a partition of the overall stock, ensuring scalability~\cite{burr06chubby}.

\section{Conclusions}
\label{sec:conclusions}

% \begin{quotation}
% \small
% \noindent\emph{We can often enhance our ability to deal with a problem by adopting a language that enables us to describe the problem in a different way.}
% \par\raggedleft--- \textup{Abelson, H. and Sussman, G.J., 1983.}\\
% \emph{Structure and Interpretation of\\ Computer Programs}~\cite{abe84sicp}
% \end{quotation}

We have presented \name, an abstraction for programming with replicated objects.
%\needsrev{While the focus is on a linguistic primitive, we build on a foundation of systems principles.}
The contribution of \name\ is twofold.
First, they \emph{decouple} an application from its underlying storage stack by drawing a clear boundary between consistency guarantees and the various methods of achieving them.
% (each method being specific to a replicated storage system).
This reduces developer effort and allows for simpler and more portable code.

Second, \name\ provide \emph{incremental consistency guarantees} (\incguarantees), which allow to compose multiple consistency levels within a single operation.
% By \incguarantees, that we provide through the \name\ abstraction, we fill a gap in the consistency/performance trade-off.
With this type of guarantees we aim to fill a gap in the consistency/performance trade-off.
Namely, applications can make last-minute decisions about what consistency level to use in an operation while this operation is executing.
This opens the door to new optimizations based on speculation or on concrete, application-specific semantics.

% Our goal was not to push the performance envelope for consistency protocols.
%% We focus instead on filling a gap: when developers choose one extreme of the consistency/performance trade-off (e.g., strong consistency) they invariably sacrifice the other variable (e.g., latency).
%% With \name, applications can obtain a low-latency response as a prelude to a strongly-consistent (high-latency) operation---or equivalently, a response with strong guarantees which arrives in the background, to confirm a speculative result of weakly-consistent operations.

We evaluated the performance and overhead of \incguarantees, as well as the impact of this novel type of guarantees on three practical systems:
(1) a microblogging service and (2) an ad serving system backed up by Cassandra, and (3) a ticket selling system based on ZooKeeper queues.
We modified both Cassandra and ZooKeeper to support \incguarantees\ with little overhead.
We showed how \incguarantees\ provided by \name\ brings substantial latency decrease for the price of small bandwidth overhead and throughput drop.

We believe that \name\ provide a new way to structure the interaction between applications and their storage by exploiting incrementality, and hence a new way to build distributed applications.

%!TEX root = ../bundle.tex

\section*{Acknowledgements}
We thank our shepherd, Timothy Roscoe, and the anonymous OSDI reviewers for their thoughtful comments which greatly improved the quality of our paper.
We are also grateful to our colleagues from the Distributed Programming Laboratory (LPD) for putting up with our recurring requests for feedback, and for the insightful discussions we had along the way with Martin Odersky (who also suggested us the name \emph{Correctables}), Edouard Bugnion, Willy Zwaenepoel, John Wilkes, Aleksandar Dragojevi\'{c}, Julia Proskurnia, Vlad Ureche, and Jad Hamza.
A special thanks goes to Kenji Relut for his help with ZooKeeper.
This work has been supported in part by the European ERC Grant 339539 - AOC and the Swiss FNS grant 20021\_147067.

% \newpage
\small

\raggedright

% Bibliography
\bibliography{bundle}
\bibliographystyle{abbrv}

\end{document}